\documentclass[letterpaper, 10 pt, conference]{ieeeconf}  

\IEEEoverridecommandlockouts                              

\usepackage{graphics} 
\usepackage{epsfig} 
\usepackage{times} 
\usepackage{amsmath} 
\usepackage{amssymb}  
\usepackage{nicematrix}

\usepackage{cite}
\usepackage{float}
\usepackage{mathrsfs}
\usepackage{mathtools}
\usepackage{svg}
\usepackage{times} 
\usepackage{subcaption}
\usepackage{comment}
\newtheorem{theorem}{Theorem}[section]

\newtheorem{proposition}[theorem]{Proposition}
\newtheorem{lemma}[theorem]{Lemma}

\newtheorem{definition}{Definition}

\newtheorem{case}{Case}

\newcommand{\PI}{{\rm{PI}}}
\newcommand{\PD}{{\rm{PD}}}
\newcommand{\uVector}{\pmb{u}}
\newcommand{\xVector}{\pmb{x}}
\newcommand{\NE}{\rm{ne}}

\newcommand{\leftparen}{\left(}

\newcommand{\rightparen}{\right)}

\title{\LARGE \bf
Conditions for Altruistic Perversity in  Two-Strategy Population Games
}

\author{Colton Hill, Philip N. Brown, and Keith Paarporn
\thanks{*This material is based upon work supported by the Air Force Office of Scientific Research under award number FA9550-23-1-0171, the National Science Foundation under award number ECCS-2013779, and the Committee for Research and Creative Works at UCCS.}
\thanks{The authors are with the University of Colorado at Colorado Springs, CO 80918, USA.
	{\tt\small \{chill13, pbrown2, kpaarpor\}@uccs.edu}}%
}

\begin{document}

\maketitle
\thispagestyle{empty}
\pagestyle{empty}

\begin{abstract}

Self-interested behavior from individuals can collectively lead to poor societal outcomes.
These outcomes can seemingly be improved through the actions of altruistic agents, which benefit other agents in the system.
However, it is known in specific contexts that altruistic agents can actually induce worse outcomes compared to a fully selfish population --- a phenomenon we term altruistic perversity.
This paper provides a holistic investigation into the necessary conditions that  give rise to  altruistic perversity.
In particular, we study the class of two-strategy population games where one sub-population is altruistic and the other is selfish.
We find that a population game can admit altruistic perversity only if the associated social welfare function is convex and the altruistic population is sufficiently large.
Our results are a first step in establishing a connection between properties of nominal agent interactions and the potential impacts from altruistic behaviors.


\end{abstract}

\section{INTRODUCTION}

In systems with a large number of interacting individuals, such as infrastructure and transportation networks, the payoff experienced by agents depends on the actions of other agents in the system.
When all agents select strategies to maximize their own payoff (commonly referred to as \textit{selfish}), it is well-known that the resulting system welfare can be suboptimal~\cite{roughgardenSelfishRoutingPrice2005,Hardin_1968}.
Whether by nature or by design, agents may also exhibit behaviors that benefit other agents in the system.
These \textit{altruistic} agents are present in several domains of study, ranging from evolutionary biology~\cite{hamiltonEvolutionAltruisticBehavior1963,lehmann2006evolution,kerr2004altruism} (e.g. the social amoeba D. discoideum~\cite{strassmann2000altruism}) and pandemic mitigation~\cite{Brown2022,dahmouni2022necessity} to the design of socio-technical systems~\cite{caragiannis2010impact,liEmployingAltruisticVehicles2021,biyikAltruisticAutonomyBeating2020,brownWhenAltruismWorse2021,hill2023tradeoff} (e.g. the use of autonomous vehicles).
Experimental research in economics has observed altruistic behavior~\cite{fehrChapterEconomicsFairness2006}, and the effects of fully-adopted altruism has been studied in a wide variety of games~\cite{chenAltruismItsImpact2014}.




Game theory offers principled approaches that have been extensively utilized to study the inefficiencies that arise from the actions of selfish agents relative to a system's optimal operation~\cite{roughgardenIntrinsicRobustnessPrice2015}.
From this perspective, a pertinent question to investigate is: in general, how does the presence of altruistic agents impact the social welfare of the system?
Indeed, social welfare is guaranteed to improve from altruistic behaviors in certain contexts -- for example, in non-atomic congestion games where all agents (at least partially) consider their impact on overall welfare~\cite{chenAltruismItsImpact2014}.

However, the benefits from altruism do not generally apply in other settings.
Counter-intuitively, it has been shown that significant negative effects can arise in systems with mixed altruistic and selfish populations.
That is, the effect of altruistic behavior can be \textit{perverse} in games with heterogeneous populations~\cite{brownCanTaxesImprove2020, sekarUncertaintyMulticommodityRouting2019}.
The potential harm caused by altruism can be quantified with the \textit{perversity index}, which measures the ratio between the social welfare in the presence of altruistic agents, with that of the social welfare that would arise if all agents behaved selfishly~\cite{brownBenefitPerversityTaxation2017}.

Much of the work done regarding perversity in games focuses specifically on the class of congestion games, where subsidies and tolls~\cite{fergusonCarrotsSticksEffectiveness2020}, choosing routes in consideration of the impact on aggregate road congestion~\cite{liEmployingAltruisticVehicles2021}, and uncertainty~\cite{sekarUncertaintyMulticommodityRouting2019} effectively measure how altruism impacts the quality of social welfare.
In series-parallel networks with arbitrary cost functions, it is known that the worst-case perversity arises when exactly half of the population is altruistic, and that the perversity increases as a function of the steepness of the cost functions~\cite{brownWhenAltruismWorse2021}.
However, altruism (even in only a fraction of the population) is guaranteed to improve social welfare in congestion games with serially-linearly-independent networks and affine cost functions, provided all agents have access to all roads~\cite{sekarUncertaintyMulticommodityRouting2019}.
Significant contributions have been made towards characterizing altruism and conditions for perversity; however, the results often come with assumptions that restrict generality.

The primary motivation of this paper is to study the emergence of altruistic perversity in general contexts that go beyond the well-studied congestion games literature.
Specifically, we use a more general context (population games) as we seek to identify conditions on the type of agent interactions
that admit welfare degradation (or improvement) in the presence of  altruistic agents.
This paper represents a first step in this direction, as we consider the impact of altruistic behavior for the entire class of $2\times 2$ population games.
This class of games encompasses a wide variety of nominal agent interactions, from Prisoner's Dilemma, Coordination, to Anti-Coordination games.

Our main result (Theorem~\ref{thm:main}) asserts that altruistic perversity can only occur if the function expressing social welfare is convex with respect to the population state.
Interestingly, perversity can occur only for a sufficiently large altruistic population.
Consequently, even all-altruistic populations have the potential to exhibit perversity.
Conversely, games with a concave welfare function cannot exhibit altruistic perversity -- the behavior of altruists can only improve societal outcomes in these cases.
We provide a detailed illustration of these phenomena in a case study of population games based on the \textit{Prisoner's Dilemma}.


\section{Model}\label{sec:model}

\subsection{Symmetric Two-Strategy Population Game with Heterogeneous Types Presented in Normal Form}

We consider a heterogeneous population consisting of a unit mass of agents, where each agent is either \textit{altruistic} or \textit{selfish}.
Altruistic agents make up mass $p_{\rm{a}}$, and selfish agents agents comprise mass $p_{\rm{s}}$, so that $p_{\rm{a}} + p_{\rm{s}} = 1$.
In symmetric two-strategy games, a $2\times2$ matrix can be used to represent the payoff of any outcome from the perspective of a row player.
Agents can either \textit{cooperate} by choosing the first row strategy, or \textit{defect} by choosing the second row strategy, and the resulting payoff depends on whether other agents cooperate or defect (the first and second column, respectively).
Thus, we write \mbox{$\mathcal{S} \coloneqq \{C, D\}$} to denote the cooperate and defect strategies available to all agents, where the payoffs are denoted by the matrix:
\begin{equation}\label{eq:A_matrix}
    A=
    \begin{bNiceMatrix}[
        first-row,code-for-first-row=\scriptstyle, 
        first-col,code-for-first-col=\scriptstyle,]
        & C & D \\
        C & R & S \\
        D & T & P
    \end{bNiceMatrix},
\end{equation}
and we may assume without loss of generality that \mbox{$R, S, T, P \in \mathbb{R}_{\geq 0}$}.
For $\tau \in \{\rm{a}, \rm{s}\}$, we write $\mathcal{X}_\tau \coloneqq \{\xVector_\tau \in \mathbb{R}^2_{\geq 0}: \sum_{i \in \mathcal{S}} x_{i,\tau} = p_\tau\}$ to denote the set of \textit{population states} for altruistic and selfish agents.
Thus $\mathcal{X} \coloneqq \mathcal{X}_{\rm{a}} \times \mathcal{X}_{\rm{s}}$ is the set of all population states, and the tuple $\xVector = (\xVector_{\rm{a}}, \xVector_{\rm{s}}) \in \mathcal{X}$ is a \textit{population state} for altruistic and selfish agents.

All agents can either cooperate or defect, so the payoff for selecting a strategy depends on how many agents of both types choose the same strategy.
Given population state $\xVector$, the \textit{utilization level} is a column vector where each entry is the sum of altruistic and selfish agents selecting the corresponding strategy in $\xVector$.
We denote the utilization level by $\uVector(\xVector): \mathcal{X} \rightarrow \mathbb{R}^2$, where $u_i(x_i) = x_{i, \rm{a}} + x_{i, \rm{s}}$ for each $i \in \mathcal{S}$.
When the context is clear, we write $\uVector$ to denote $\uVector(\xVector)$.
Since there are two strategies, we may represent $\uVector$ by $u \in \mathbb{R}$, where $\uVector = \begin{bmatrix} u & 1-u \end{bmatrix}^\top$. Here, $u$ is the fraction of agents cooperating, and $1-u$ is the fraction of agents who defect.

We consider the set of altruistic and selfish populations and their strategies as established and identify a game with the payoffs experienced by agents for their decisions.
The \textit{payoff function} is a continuous mapping that associates the utilization level for a population state with a payoff vector:
\begin{align}\label{eq:payoff_function}
    f(\uVector(\xVector)) &: \mathcal{X} \rightarrow \mathbb{R}^2.
\end{align}
Since the payoffs agents receive is based on the matrix defined by~\eqref{eq:A_matrix}, we write $f(\uVector) \coloneqq A\uVector$.
We then measure the \textit{total social welfare}, given a population state $\xVector$, by
\begin{align}\label{eq:Social_Welfare}
     \nonumber
    W(\uVector) &\coloneqq
    \uVector^{\top} A \uVector
    \\ &=
    (R \hspace{-0.3mm} + \hspace{-0.3mm} P \hspace{-0.3mm} - \hspace{-0.3mm} (S \hspace{-0.3mm} + \hspace{-0.3mm} T))u^2 \hspace{-0.3mm} + \hspace{-0.3mm} 
    (S \hspace{-0.3mm} +\hspace{-0.3mm} T \hspace{-0.3mm} - \hspace{-0.3mm} 2P)u \hspace{-0.3mm} + 
    \hspace{-0.3mm} P,
\end{align}
where $\uVector^\top$ is the transpose of the utilization level for $\xVector$.
The payoffs experienced by agents is determined by their type.
Selfish agents are concerned only with maximizing their own payoff, so they aim to select the strategy that maximizes the actual payoff for a given utilization level $\uVector$:
\begin{align}\label{eq:selfish_payoff}
    \nonumber
    f_{\rm{s}}(\uVector) &\coloneqq 
    A\uVector 
    \\ \nonumber &=
    \begin{bmatrix}
        Ru + S(1-u) \\
        Tu + P(1-u)
    \end{bmatrix}
    \\ &=
    \begin{bmatrix}
        f_{C, \rm{s}}(u) \\
        f_{D, \rm{s}}(u)
    \end{bmatrix}.
\end{align}
In contrast, altruistic agents are concerned with increasing social welfare. 
Since each agent is infinitesimal, and there are only two strategies to choose from, they select the strategy that is in the direction of increased social welfare.
The gradient of the social welfare function~\eqref{eq:Social_Welfare}, projected onto the unit simplex, represents the desired payoff for altruists:
\begin{align}\label{eq:altruistic_payoff}
    \nonumber
    f_{\rm{a}}(\uVector) &\coloneqq \nabla_{\uVector}W(\uVector)
    \\ \nonumber &=
    \begin{bmatrix}
        (2R - (S+T))u + (S+T - 2P)(1-u) \\
        (S+T - 2R)u + (2P - (S+T))(1-u) 
    \end{bmatrix} 
    \\ &=
    \begin{bmatrix}
        f_{C, \rm{a}}(u) \\
        f_{D, \rm{a}}(u)
    \end{bmatrix},
\end{align}
where $f_{C, \rm{a}}(u) = -f_{D, \rm{a}}(u)$.
An instance of a population game with selfish and altruistic types is fully specified by the tuple $G = \leftparen \mathcal{S}, f_{\tau \in \{\rm{a}, \rm{s}\}}, p_{\rm{a}}\rightparen$.

A standard solution concept for population games is the Nash equilibrium, which describes a state in which no agent can benefit from unilaterally changing their strategy.
\begin{definition} A \textit{Nash equilibrium} is a population state \mbox{$\xVector \in \mathcal{X}$} such that for each type $\tau \in \{\rm{a}, \rm{s}\}$:
\begin{align}\label{eq:Nash_Eq_Cond}
     x_{i, \tau} > 0 &\Longrightarrow f_{i,\tau}(\uVector(\xVector)) \geq f_{i^{\prime},\tau}(\uVector(\xVector)) \hspace{2mm} \forall i, i^{\prime} \in \mathcal{S},
\end{align}
\end{definition}
\vspace{2mm}
where a population state corresponding to a Nash equilibrium is denoted $\xVector^{\NE} = (\xVector^{\NE}_{\rm{a}}, \xVector^{\NE}_{\rm{s}})$.
For each $\tau \in \{\rm{a}, \rm{s}\}$, we may represent $\xVector_\tau$ by $x_\tau \in \mathbb{R}$, since $\xVector_\tau = \begin{bmatrix} x_\tau & p_\tau-x_\tau \end{bmatrix}^\top$, where $x_\tau$ is the fraction of agents cooperating, and $p_\tau-x_\tau$ is the fraction of agents defecting.
The utilization level that corresponds to a Nash equilibrium, $\uVector(\xVector^{\NE})$, is often denoted $\uVector^{\NE}$ (or simply $u^{\NE}$).
Since only two strategies are available, all Nash equilibria must satisfy one of the following conditions:
\begin{align}\label{eq:equilibrium_conditions}
    \nonumber
    x^{\NE}_{\tau} = 0 &\Longleftrightarrow f_{C, \tau}(u^{\NE}) < f_{D, \tau}(u^{\NE}),
    \\
    x^{\NE}_{\tau} \in (0, p_\tau) &\Longleftrightarrow f_{C, \tau}(u^{\NE}) = f_{D, \tau}(u^{\NE}),
    \\ \nonumber
    x^{\NE}_{\tau} = p_{\tau} &\Longleftrightarrow f_{C, \tau}(u^{\NE}) > f_{D, \tau}(u^{\NE}).
\end{align}
The linearity of the payoff functions implies that there is only one Nash equilibrium $u^*_{\tau} \in [0,1]$ for each $\tau \in \{\rm{a}, \rm{s}\}$ such that $f_{C, \tau}(u^*_{\tau}) = f_{D, \tau}(u^*_{\tau})$:
\begin{equation}\label{eq:u_s_star}
    u^*_{\rm{s}} \coloneqq \frac{P-S}{R+P-(S+T)},
\end{equation}
and
\begin{equation}\label{eq:u_a_star}
    u^*_{\rm{a}} \coloneqq \frac{2P-(S+T)}{2(R+P-(S+T))}.
\end{equation}
The case that $f_{C, \tau}(u) = f_{D, \tau}(u)$ for all $u \in [0,1]$ is trivial, since it implies $W(\uVector)$ is constant.
For a game \mbox{$G=\leftparen \mathcal{S}, f_{\tau \in \{\rm{a}, \rm{s}\}}, p_{\rm{a}}\rightparen$}, we write the set of population states that result in a  Nash equilibrium for all agents as \mbox{$\mathcal{X}^{\NE}(G) \subseteq \mathcal{X}$}.
The set of Nash equilibria for an all-selfish version of $G$ is denoted
$\mathcal{X}^{\NE}_{\rm{s}}(G) \coloneqq {\cal X}^{\NE} \leftparen \mathcal{S}, f_{\tau \in \{\rm{a}, \rm{s}\}}, 0 \rightparen $, and the corresponding set for an all-altruistic version of $G$ is denoted $\mathcal{X}^{\NE}_{\rm{a}}(G) \coloneqq {\cal X}^{\NE}\leftparen \mathcal{S}, f_{\tau \in \{\rm{a}, \rm{s}\}}, 1 \rightparen$.
We often write ${\cal X}^{\NE}(G)$, $\mathcal{X}^{\NE}_{\rm{s}}(G)$, and $\mathcal{X}^{\NE}_{\rm{a}}(G)$ as ${\cal X}^{\NE}$, $\mathcal{X}^{\NE}_{\rm{s}}$, and $\mathcal{X}^{\NE}_{\rm{a}}$ (respectively) when the dependence on $G$ is clear.


\subsection{Performance Metric: Perversity Index}

In this paper, we study the \textit{perversity index}~\cite{brownCanTaxesImprove2020} to understand the effects of heterogeneous altruism.
The perversity index captures the potential negative impact the presence of altruism has in a population game, relative to its all-selfish counterpart (i.e. $p_{\rm{a}} = 0$).
The perversity index is defined as the worst-case ratio of the social welfare of a heterogeneous Nash equilibrium with that of the social welfare that arises from an all-selfish Nash equilibrium:
\begin{equation}\label{eq:PI}
    \PI(G) \coloneqq
    \frac
    {\min_{\xVector \in \mathcal{X}^{\NE}(G)} W(\uVector(\xVector))}
    {\max_{\xVector \in \mathcal{X}^{\NE}_{\rm{s}}(G)} W(\uVector(\xVector))}.
\end{equation}
$\PI(G) < 1$ indicates the presence of altruists can hurt social welfare at equilibrium -- here, we say that the game exhibits altruistic perversity.
Likewise, \mbox{$\PI(G) > 1$} indicates the presence of altruists improves social welfare at equilibrium.

\section{Classifying Games with Perversity in Symmetric Two-Strategy Population Games}\label{sec:result}

One might expect that introducing altruistic agents in a population game would lead to Nash equilibria with improved social welfare.
We show that this need not be the case.
Indeed, we seek to identify necessary conditions on the underlying population game, specifically the payoff functions and welfare, that admit worsened social welfare in the presence of altruists compared to an all-selfish population.
That is, we seek to classify games $G$ that admit altruistic perversity, $\PI(G) < 1$.
Our main result is given below.

\begin{theorem}\label{thm:main}
    Let $G$ be a heterogeneous symmetric two-strategy population game.
    If the presence of altruistic agents in $G$ admits altruistic perversity, i.e. $\PI(G) < 1$, then the welfare function defined by~\eqref{eq:Social_Welfare} is convex.
\end{theorem}

An example of the altruistic perversity characterized by this result is presented in Section~\ref{sec:prisoners_dilemma}.
The proof (completed via the contrapositive result) of Theorem~\ref{thm:main} is presented in section~\ref{sec:proof}.
The contrapositive states that if the welfare function is concave, then the perversity index is greater than $1$.
This implies the presence of altruists cannot degrade equilibrium welfare in games with concave welfare functions.
Since altruists choose actions in the direction of the welfare gradient, they act as a local gradient ascent on $W$, so a sufficient amount of altruists will lead to welfare maximization.

Conversely, when $W$ is convex, social welfare is maximized at an extreme point where all agents play the same action.
In this case, altruists can still increase welfare, and are at equilibrium when $W$ is maximized.
However, altruists now have the potential to induce deteriorated welfare because the local minimizer of $W$ coincides with a Nash equilibrium for altruists.
In particular, if $u^*_{\rm{a}}$ is feasible and exceeds how many selfish agents cooperate, and the altruistic population exceeds $u^*_{\rm{a}}$, the worst-case welfare is achieved since $u^*_{\rm{a}}$ coincides with the global minimum of $W$.
Counter-intuitively, this means that perverse outcomes do not emerge unless there is a sufficiently large population of altruists.

In the next section, we concretely illustrate the altruistic perversity that emerges when the underlying population game is a Prisoner's Dilemma.


\begin{figure*}[t]
    \centering
        \begin{subfigure}[t]{.33\textwidth}
            \includegraphics[width=\textwidth]{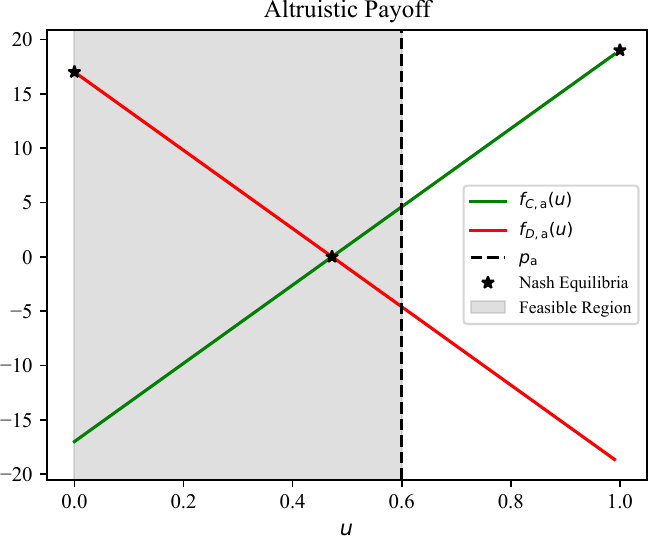}
            \caption{}
            \label{subfig:alt_payoff}
        \end{subfigure}\hfill
        \begin{subfigure}[t]{.33\textwidth}
            \includegraphics[width=\textwidth]{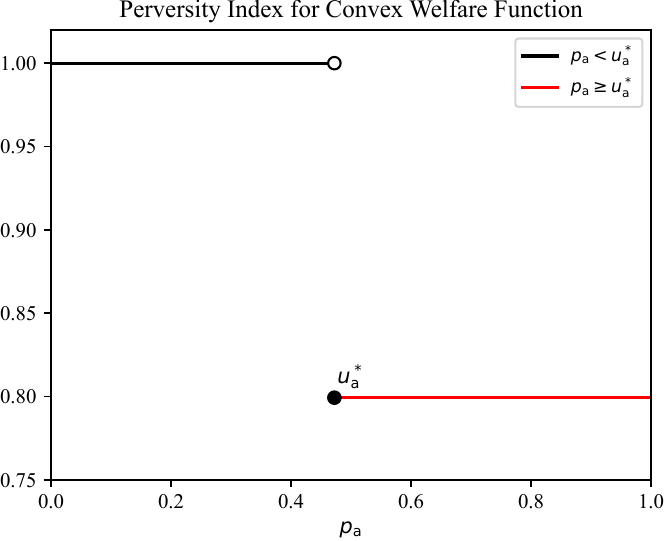}
            \caption{}
            \label{subfig:convex_PI}
        \end{subfigure}\hfill
        \begin{subfigure}[t]{.33\textwidth}
            \includegraphics[width=\textwidth]{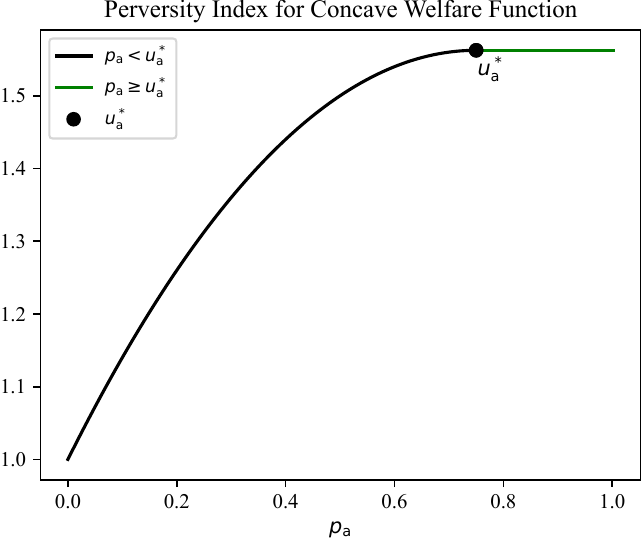}
            \caption{}
            \label{subfig:concave_PI}
        \end{subfigure}
        \caption{Fig.~\ref{subfig:alt_payoff} characterizes the payoff functions and possible Nash equilibria for altruists in an example game where the welfare function is convex: $R = 21$, $S = 1$, $T = 22$, $P = 20$.
        The stars represent the Nash equilibria available to altruists when $p_{\rm{a}} = 1$, and the shaded area contains feasible sub-population states for a given altruistic population.
        Fig.~\ref{subfig:convex_PI} represents the perversity index as a function of the altruistic population, $\PI(p_a)$, for the same example game.
        Here, the perversity index is a piecewise constant function since, if their population is too small, altruists choose to defect just like selfish agents.
        If their population exceeds $u_{\rm{a}}^*$, altruists may choose the mixed Nash equilibrium, which results in the worst-case welfare.
        In this example, altruistic perversity can significantly degrade welfare, resulting in a $20\%$ drop in performance.
        Fig.~\ref{subfig:concave_PI} represents $\PI(p_a)$ for an example game where the welfare function is concave: $R = 3$, $S = 1$, $T = 6$, $P = 2$.
        Here, the perversity index is continuous because the behavior of the altruistic payoffs is unlike that of Fig.~\ref{subfig:alt_payoff}; altruists cooperate until the population is large enough to choose the mixed Nash equilibrium, resulting in the best-case welfare.}
        \label{PD_figs}
        \vspace{-2mm}
\end{figure*}

\section{Case Study: Prisoner's Dilemma}\label{sec:prisoners_dilemma}

Here, we present the \textit{Prisoner's Dilemma} population game as an example of the perversity that can arise as described in Theorem~\ref{thm:main}.
Suppose the entries in the payoff matrix defined by~\eqref{eq:A_matrix} satisfy $S \hspace{-0.35mm} < \hspace{-0.35mm} P \hspace{-0.35mm} < \hspace{-0.35mm} R \hspace{-0.35mm} < \hspace{-0.35mm} T$, then the symmetric two-strategy population game becomes a Prisoner's Dilemma, which we denote $\PD(p_{\rm{a}})$.
In the all-selfish version of $\PD(p_{\rm{a}})$ (i.e. $p_{\rm{a}} = 0$), the only Nash equilibrium is when all agents defect, i.e. $u^{\NE}_{\rm{s}} = 0$.
Thus, all agents get the punishment payoff, and the social welfare is \mbox{$W(0) = P$}.
So, the perversity index defined by~\eqref{eq:PI} becomes
\begin{equation}\label{eq:PD_PI}
    \PI(\PD(p_{\rm{a}})) = \min_{\xVector \in \mathcal{X}^{\NE}(G)} \frac{W(\uVector^{\NE})}{P}.
\end{equation}


The result below fully characterizes~\eqref{eq:PD_PI} as a function of the altruistic mass, $p_{\rm{a}}$.


\begin{proposition}\label{prop:PD_characterization}
    Let $\PD(p_{\rm{a}})$ be a heterogeneous symmetric two-strategy Prisoner's Dilemma population game, where $p_{\rm{a}}$ is the mass of altruistic agents.
    Define $\delta \coloneqq R+P-(S+T)$ and \mbox{$\beta \coloneqq S+T-2P$}.
    If $W(u)$ is convex, then the perversity index is given by:
    \begin{equation}\label{eq:PD_convex_PI_characterization}
        \PI(\PD(p_{\rm{a}})) = 
        \begin{cases}
            1, & \text{if } p_{\rm{a}} < u^*_{\rm{a}} \\
            1 - \frac{\beta^2}{4 P \delta}, & \text{if } p_{\rm{a}} \geq u^*_{\rm{a}}
        \end{cases},
    \end{equation}
    and if $W(u)$ is concave, then the perversity index is given by:
        \begin{equation}\label{eq:PD_concave_PI_characterization}
        \PI(\PD(p_{\rm{a}})) = 
        \begin{cases}
            \frac{\delta p^2_{\rm{a}} + \beta p_{\rm{a}}}{P} + 1, & \text{if } p_{\rm{a}} < u^*_{\rm{a}}\\
            1 - \frac{\beta^2}{4 P \delta}, & \text{if } p_{\rm{a}} \geq u^*_{\rm{a}}
        \end{cases}.
    \end{equation}
    \vspace{0.5mm}
\end{proposition}

The proof is presented in the Appendix, but a short discussion is presented here to describe equilibria and payoffs to agents in this type of game.
Here, selfish agents defect regardless of whether altruists cooperate or defect, emphasizing the intuition behind Theorem~\ref{thm:main} that the improvement (or degradation) of social welfare is dependent on the choices altruists make.
Since selfish agents always defect, and since $P<R$, the payoff function defined by~\eqref{eq:altruistic_payoff} that altruists use is accurately informing them of where the locally maximum welfare is as expected.
However, if welfare is convex, a flaw arises because altruists are indifferent about cooperating or defecting at the global minimum for welfare, because the payoff at this point is $0$ regardless of the strategy altruists select.
Fig.~\ref{subfig:alt_payoff} depicts the payoffs to altruists and Fig.~\ref{subfig:convex_PI} depicts the altruistic perversity that arises.
The case that welfare is concave does not suffer from this issue, as the point at which altruists are indifferent is actually the global maximum for welfare, and it is the only point at which each altruist is content with their decision.
The perversity index in this case is depicted in Fig.~\ref{subfig:concave_PI}.

\section{Proof of Theorem~\ref{thm:main}}\label{sec:proof}

We first provide a brief outline of the proof, then present a lemma and discuss its importance.
The proof is accomplished by showing the contrapositive: if the welfare function defined by~\eqref{eq:Social_Welfare} is strictly concave, then $\PI(G) \geq 1$.
The contrapositive is proved with the following cases where $u^*_{\rm{a}}$ is defined by~\eqref{eq:u_a_star}:
\begin{itemize}
    \item Case~\ref{case:1}: if $u^*_{\rm{a}} \leq 0$, then $\PI(G) \geq 1$.
    \item Case~\ref{case:2}: if $u^*_{\rm{a}} \geq 1$, then $\PI(G) \geq 1$.
    \item Case~\ref{case:3}: if $u^*_{\rm{a}} \in (0,1)$ and \mbox{$u^*_{\rm{a}} \leq p_{\rm{a}}$}, then $\PI(G) \geq 1$.
    \item Case~\ref{case:4}: if $u^*_{\rm{a}} \in (0,1)$ such that $u^*_{\rm{a}} > p_{\rm{a}}$, and \mbox{$\mathcal{X}^{\NE}_{\rm{s}} \subseteq \{0,1\}$}, then $\PI(G) \geq 1$.
    \item Case~\ref{case:5}: if $u^*_{\rm{a}} \in (0,1)$ such that $u^*_{\rm{a}} > p_{\rm{a}}$, and \mbox{$\mathcal{X}^{\NE}_{\rm{s}} = \{u^*_{\rm{s}}\}$}, then $\PI(G) \geq 1$.
\end{itemize}

Our sole lemma characterizes potential selfish and altruistic Nash equilibria under concave social welfare functions.
\begin{lemma}\label{lem:X_tau_cardinality}
    Let $G$ be a symmetric two-strategy population game.
    If $W(u)$ is strictly concave, then $|\mathcal{X}^{\NE}_{\rm{s}}| = 1$, and $|\mathcal{X}^{\NE}_{\rm{a}}| = 1$.
\end{lemma}

The proof of Lemma~\ref{lem:X_tau_cardinality} appears in the Appendix.
The implication is that each population of agents has only one Nash equilibrium that they are trying to reach.
Thus, the lemma is useful in showing that, in heterogeneous games, agents of each type still only have one Nash equilibrium.
Intuitively, this means that each altruistic and selfish agent is making decisions with limited regard to what others are doing.
We proceed with the proof of the main result:

\textit{Proof of Theorem~\ref{thm:main}:}
Let $\uVector^{\NE}_{\rm{s}}$ and $\uVector^{\NE}$ be the utilization level for an all-selfish Nash equilibrium and a heterogeneous Nash equilibrium, respectively.
We use \mbox{$\xVector^{\NE} = \begin{bmatrix} (x^{\NE}_{\rm{a}}, x^{\NE}_{\rm{s}}) & (p_{\rm{a}} - x^{\NE}_{\rm{a}}, p_{\rm{s}} - x^{\NE}_{\rm{s}})\end{bmatrix}$}, to denote a heterogeneous Nash equilibrium
where $ x^{\NE}_{\rm{a}}, x^{\NE}_{\rm{s}} \in [0,1]$.
Hence, the heterogeneous utilization level is
\begin{align}
    \nonumber
    \uVector^{\NE} &= \begin{bmatrix}u^{\NE} & 1 - u^{\NE}\end{bmatrix}^\top
    \\ \nonumber &= 
    \begin{bmatrix}x^{\NE}_{\rm{a}} + x^{\NE}_{\rm{s}} & 1 - (x^{\NE}_{\rm{a}} + x^{\NE}_{\rm{s}})\end{bmatrix}^\top.
\end{align}
Since $W(u)$ is strictly concave, it is known that $u^*_{\rm{a}}$ is the global maximum, so $W(u^*_{\rm{a}}) \geq W(u)$ for all $u$.
The following cases complete the proof; we provide intuition here, and proceed with the proof of each case in the appendix.

\begin{case}\label{case:1}
    If $u^*_{\rm{a}} \leq 0$, then $\PI(G) \geq 1$.
\end{case}

Case~\ref{case:1} implies that altruists always choose to defect in a game where the social welfare function is decreasing from $0$ to $1$.
Hence, the number of agents cooperating will always be less in the heterogeneous game than in the all-selfish game.

\begin{case}\label{case:2}
    If $u^*_{\rm{a}} \geq 1$, then $\PI(G) \geq 1$.
\end{case}

Case~\ref{case:2} implies that altruists always cooperate in a game where the social welfare function is increasing from $0$ to $1$.
Thus, the number of agents cooperating will always be higher in the heterogeneous game than in the all-selfish game.

\begin{case}\label{case:3}
    If $u^*_{\rm{a}} \in (0,1)$ such that $u^*_{\rm{a}} \leq p_{\rm{a}}$, then \mbox{$\PI(G) \geq 1$}.
\end{case}

Case~\ref{case:3} shows that if the mass of altruistic agents is large enough, they are guaranteed to move social welfare closer to their preferred mixed Nash equilibrium, the maximum social welfare, than selfish agents would do on their own.

\begin{case}\label{case:4}
     If $u^*_{\rm{a}} \in (0,1)$ such that $u^*_{\rm{a}} > p_{\rm{a}}$, and \mbox{$\mathcal{X}^{\NE}_{\rm{s}} \subseteq \{1,0\}$}, then $\PI(G) \geq 1$.
\end{case}

Case~\ref{case:4} implies that even a relatively small population of altruists is able to improve the overall social welfare, regardless of whether the selfish agents cooperate or defect.

\begin{case}\label{case:5}
    If $u^*_{\rm{a}} \in (0,1)$ such that $u^*_{\rm{a}} > p_{\rm{a}}$, and \mbox{$\mathcal{X}^{\NE}_{\rm{s}} = \{u^*_{\rm{s}}\}$}, then $\PI(G) \geq 1$.
\end{case}

Case~\ref{case:5} also implies that a relatively small population of altruists is able to improve overall social welfare, in the instance that selfish agents now prefer a mixed Nash equilibrium.

Cases~\ref{case:1}-\ref{case:5} show that if $W$ is strictly concave, then \mbox{$\PI(G) \geq 1$} for any value obtained by $u^*_{\rm{a}} \in \mathbb{R}$.
Hence the contrapositive is shown: if $\PI(G) < 1$, then $W$ is convex. \hfill $\blacksquare$

\section{CONCLUSIONS}

We have provided general conditions for when the presence of altruistic agents can actually worsen social welfare in the class of two-strategy population games.
These results are an initial step to identifying how the structure of agent interactions in a population may dictate whether altruistic behavior improves or degrades social welfare.
Future work warrants the investigation into even more generalized relationships between agents and social welfare. 
Arbitrary $n$-strategy population games with $m$-population types, where each population uniquely weighs how much it maximizes its own payoff versus social welfare, as well as stable outcomes associated with evolutionary dynamics will be studied.




\section*{APPENDIX}
$W(u)$ being strictly concave has the following implication, which is stated here for convenience:
\begin{equation}\label{eq:W_concave_implication}
    R+P-(S+T) < 0,
\end{equation}

\noindent First, we include the proof of Lemma~\ref{lem:X_tau_cardinality}.

\textit{Proof of Lemma~\ref{lem:X_tau_cardinality}:}
Suppose to the contrary that the claim is false.
Since the payoff functions for agents of both types is affine, the possible cardinality of $\mathcal{X}^{\NE}_{\rm{s}}$ and $\mathcal{X}^{\NE}_{\rm{a}}$ is $1$, $3$, or $\infty$.
If the cardinality of $\mathcal{X}^{\NE}_{\rm{s}}$ or $\mathcal{X}^{\NE}_{\rm{a}}$ is $\infty$, the implication is that the welfare function is constant, so we may assume $\mathcal{X}^{\NE}_{\rm{s}} = \{1, u^*_{\rm{s}}, 0\}$, or $\mathcal{X}^{\NE}_{\rm{a}} = \{1, u^*_{\rm{a}}, 0\}$.
Suppose first that $\mathcal{X}^{\NE}_{\rm{s}} = \{1, u^*_{\rm{s}}, 0\}$.
Since $1$ is a Nash equilibrium, $f_{D, \rm{s}}(1) \leq f_{C, \rm{s}}(1)$, i.e. $T \leq R$.
Similarly, since $0$ is a Nash equilibrium, $f_{C, \rm{s}}(0) \leq f_{D, \rm{s}}(0)$, i.e. $S \leq P$.
Thus $S+T \leq P+T \leq P+R$, i.e. $R+P - (S+T) \geq 0$, contradicting~\eqref{eq:W_concave_implication}.
Now suppose $\mathcal{X}^{\NE}_{\rm{a}} = \{1, u^*_{\rm{a}}, 0\}$.
Since $1$ is a Nash equilibrium, $f_{D, \rm{a}}(1) \leq f_{C, \rm{a}}(1)$, i.e. $S+T-2R \leq 2R-(S+T)$.
Similarly, since $0$ is a Nash equilibrium, \mbox{$f_{C, \rm{a}}(0) \leq f_{D, \rm{a}}(0)$}, i.e. \mbox{$S+T-2P \leq 2P-(S+T)$}.
But then $ 0\leq 2(R+P-(S+T))$, contradicting~\eqref{eq:W_concave_implication}.\hfill $\blacksquare$

\subsection*{Proof of Cases~\ref{case:1}-\ref{case:5} for Theorem~\ref{thm:main}}

\textit{Proof of Case~\ref{case:1}:}
Since $u^*_{\rm{a}} \leq 0$ is the global maximum, we have that $u^*_{\rm{a}} \leq u$, and thus $W(u) \leq W(u^*_{\rm{a}})$ for all $u \in [0,1]$.
Then $2P - (S+T) \geq 0$ implies \mbox{$f_{C, \rm{a}}(0) = S+T-2P \leq 0$}.
Now,~\eqref{eq:W_concave_implication} implies $f_{C, \rm{a}}(u)$ is decreasing, so $f_{C,\rm{a}}(u) \leq 0$ for all $u$.
Since $f_{D,\rm{a}}(u) = -f_{C,\rm{a}}(u)$, \mbox{$x^{\NE}_{\rm{a}} = 0$} is the only Nash equilibrium for altruists.
So, for any $p_{\rm{a}} \in [0,1]$, we have that altruists always defect, so \mbox{$u^{\NE} \leq u^{\NE}_{\rm{s}}$}.
Thus $W(u^{\NE}_{\rm{s}}) \leq W(u^{\NE})$, since $W(u)$ is decreasing for all \mbox{$u \in[0,1]$}. \hfill $\blacksquare$

\textit{Proof of Case~\ref{case:2}:}
Since $u^*_{\rm{a}} \geq 1$ is the global maximum, we have that $u \leq u^*_{\rm{a}}$, and thus $W(u) \leq W(u^*_{\rm{a}})$ for all $u \in [0,1]$.
Also, $2R-(S+T) \geq 0$ implies \mbox{$f_{C, \rm{a}}(1) = 2R-(S+T) \geq 0$}.
Since $u^*_{\rm{a}} > 0$, and by~\eqref{eq:W_concave_implication}, it is the case that $f_{C, \rm{a}}(0) = S+T-2P \geq 0$.
Hence $f_{C, \rm{a}}(u) \geq 0$ for all $u$.
Since \mbox{$f_{D, \rm{a}}(u) = -f_{C, \rm{a}}(u)$}, \mbox{$x^{\NE}_{\rm{a}} = p_{\rm{a}}$} is the only Nash equilibrium for altruists.
So, for any $p_{\rm{a}} \in [0,1]$, we have that altruists always cooperate, so that $u^{\NE}_{\rm{s}} \leq u^{\NE}$.
Hence, $W(u^{\NE}_{\rm{s}}) \leq W(u^{\NE})$. \hfill $\blacksquare$

\textit{Proof of Case~\ref{case:3}:}
Since $u^*_{\rm{a}} \in (0, 1)$, \mbox{$\mathcal{X}^{\NE}_{\rm{a}} = \{u^*_{\rm{a}}\}$} by Lemma~\ref{lem:X_tau_cardinality}.
If $x^{\NE}_{\rm{s}} \leq u^*_{\rm{a}}$, then we claim $x^{\NE}_{\rm{a}} = u^*_{\rm{a}} - x^{\NE}_{\rm{s}}$.
We can see that $x^{\NE}_{\rm{a}}$ is feasible since $0 \leq x^{\NE}_{\rm{a}} \leq u^*_{\rm{a}} \leq p_{\rm{a}}$, and 
\begin{align}
    \nonumber
    f_{C, \rm{a}}(x^{\NE}_{\rm{a}} + x^{\NE}_{\rm{s}}) &=
    f_{C, {\rm{a}}}(u^*_{\rm{a}})
    \\ \nonumber &=
    f_{D, {\rm{a}}}(u^*_{\rm{a}}).
\end{align}
Further, this is the only Nash equilibrium by Lemma~\ref{lem:X_tau_cardinality}, thus $W(u^{\NE}_{\rm{s}}) \leq W(u^{\NE}) = W(u^*_{\rm{a}})$.
If $x^{\NE}_{\rm{s}} > u^*_{\rm{a}}$, then we claim $x^{\NE}_{\rm{a}} = 0$; to be clear, this implies altruists defect (choose $f_{D, \rm{a}}(u)$).
It is clear that $x^{\NE}_{\rm{a}}$ is feasible, and since \mbox{$\mathcal{X}^{\NE}_{\rm{a}} = \{u^*_{\rm{a}}\}$} and $u^*_{\rm{a}} < x^{\NE}_{\rm{s}}$, we have that 
\begin{align}
    \nonumber
    f_{C, \rm{a}} (u^{\NE}) &\leq
    f_{C, \rm{a}} (u^*_{\rm{a}}) 
    \\ \nonumber &=
    f_{D, \rm{a}} (u^*_{\rm{a}})
    \\ \nonumber &\leq
    f_{D, \rm{a}} (u^{\NE}).
\end{align}
Thus, $x^{\NE}_{\rm{a}} = 0$ is the only Nash equilibrium by Lemma~\ref{lem:X_tau_cardinality}, and it follows that $u^{\NE} \leq u^{\NE}_{\rm{s}}$.
Hence, we have that $W(u^{\NE}_{\rm{s}}) \leq W(u^{\NE})$ since $W(u)$ is decreasing for $u^*_{\rm{a}} \leq u$. \hfill $\blacksquare$

\textit{Proof of Case~\ref{case:4}:}
By Lemma~\ref{lem:X_tau_cardinality}, $\mathcal{X}^{\NE}_{\rm{a}} = \{u^*_{\rm{a}}\}$ since $u^*_{\rm{a}} \in (0,1)$, and $\mathcal{X}^{\NE}_{\rm{s}}$ is equal to the set containing only one element of $\{1, 0\}$.
If $\mathcal{X}^{\NE}_{\rm{s}} = \{1\}$, then $u^{\NE}_{\rm{s}} = 1$, and $f_{C, \rm{s}}(1) \geq f_{D, \rm{s}}(1)$, i.e. $R \geq T$.
Thus, by~\eqref{eq:W_concave_implication}, it must also be the case that $f_{C, \rm{s}}(0) = S \geq P = f_{D, \rm{s}}(0)$.
Hence, $f_{C, \rm{s}}(u) \geq f_{D, \rm{s}}(u)$ for all $u$, and so $x^{\NE}_{\rm{s}} = p_{\rm{s}}$ is the only heterogeneous Nash equilibrium for selfish agents.
Now, $u^{\NE} \leq u^{\NE}_{\rm{s}}$, and since $\mathcal{X}^{\NE}_{\rm{a}} = \{u^*_{\rm{a}}\}$, we have that $u^*_{\rm{a}} \leq u^{\NE} \leq u^{\NE}_{\rm{s}} = 1$, so that $W(u^{\NE}_{\rm{s}}) \leq W(u^{\NE})$.
If $\mathcal{X}^{\NE}_{\rm{s}} = \{0\}$, then $u^{\NE}_{\rm{s}} = 0$, so $f_{D, \rm{s}}(0) \geq f_{C, \rm{s}}(0)$, i.e. $P \geq S$.
Thus, by~\eqref{eq:W_concave_implication}, it must also be the case that $f_{D, \rm{s}}(1) = T \geq R = f_{C, \rm{s}}(1)$.
Thus $f_{D, \rm{s}}(u) \geq f_{C, \rm{s}}(u)$ for all $u$, and so $x^{\NE}_{\rm{s}} = 0$ is the only heterogeneous Nash equilibrium for selfish agents.
It is clear that $u^{\NE}_{\rm{s}} \leq u^{\NE}$, and since $p_{\rm{a}} < u^*_{\rm{a}}$, it follows that $0 = u^{\NE}_{\rm{s}} \leq u^{\NE} \leq u^*_{\rm{a}}$.
Thus $W(u^{\NE}_{\rm{s}}) \leq W(u^{\NE})$. \hfill $\blacksquare$

\textit{Proof of Case~\ref{case:5}:}
If $x^{\NE}_{\rm{a}} \leq u^*_{\rm{s}}$ and $u^*_{\rm{s}} - x^{\NE}_{\rm{a}} \leq p_{\rm{s}}$, then we claim that $x^{\NE}_{\rm{s}} = u^*_{\rm{s}} - x^{\NE}_{\rm{a}}$.
It is clear that $x^{\NE}_{\rm{s}}$ is feasible since $0 \leq x^{\NE}_{\rm{s}} \leq u^*_{\rm{s}} \leq p_{\rm{s}}$, and 
\begin{align}
    \nonumber
    f_{C, \rm{s}}(u^{\NE}) &= 
    f_{C, \rm{s}}(u^*_{\rm{s}})
    \\ \nonumber &=
    f_{D, \rm{s}}(u^*_{\rm{s}}).
\end{align}
Note that this is the only Nash equilibrium by Lemma~\ref{lem:X_tau_cardinality}.
Thus, $W(u^{\NE}) = W(u^*_{\rm{s}}) = W(u^{\NE}_{\rm{s}})$, i.e. $W(u^{\NE}_{\rm{s}}) \leq W(u^{\NE})$ trivially.
If $x^{\NE}_{\rm{a}} \leq u^*_{\rm{s}}$ and $u^*_{\rm{s}} - x^{\NE}_{\rm{a}} > p_{\rm{s}}$, then we claim that \mbox{$x^{\NE}_{\rm{s}} = p_{\rm{s}}$}.
It is clear that $x^{\NE}_{\rm{s}}$ is feasible, and 
\begin{align}
    \nonumber
    f_{D, \rm{s}}(x^{\NE}_{\rm{s}} + x^{\NE}_{\rm{a}})
    &\leq
    f_{D, \rm{s}}(u^*_{\rm{s}} - x^{\NE}_{\rm{a}} + x^{\NE}_{\rm{a}})
    \\ \nonumber &=
    f_{D, \rm{s}}(u^*_{\rm{s}})
    \\ \nonumber &=
    f_{C, \rm{s}}(u^*_{\rm{s}})
    \\ \nonumber &\leq
    f_{C, \rm{s}}(p_{\rm{s}} + x^{\NE}_{\rm{a}})
    \\ \nonumber &=
    f_{C, \rm{s}}(u^{\NE}).
\end{align}
This is also the only Nash equilibrium by Lemma~\ref{lem:X_tau_cardinality}.
Now, $u^{\NE} = x^{\NE}_{\rm{s}} + x^{\NE}_{\rm{a}} < u^*_{\rm{s}} - x^{\NE}_{\rm{a}} + x^{\NE}_{\rm{a}} = u^*_{\rm{s}} = u^{\NE}_{\rm{s}}$, i.e. \mbox{$u^{\NE} < u^{\NE}_{\rm{s}}$}.
Also, $u^{\NE} = p_{\rm{s}} + x^{\NE}_{\rm{a}} \geq u^*_{\rm{a}}$ (otherwise, altruists are not at Nash equilibrium or $u^*_{\rm{a}} \geq 1$, both contradictions).
Hence $u^*_{\rm{a}} \leq u^{\NE} \leq u^{\NE}_{\rm{s}}$, so that $W(u^{\NE}_{\rm{s}}) \leq W(u^{\NE})$.
If $x^{\NE}_{\rm{a}} > u^*_{\rm{s}}$, then we claim $x^{\NE}_{\rm{s}} = 0$ (selfish agents defect and choose $f_{D, \rm{s}}(u)$).
It is clear that $x^{\NE}_{\rm{s}}$ is feasible, and since $\mathcal{X}^{\NE}_{\rm{s}} = \{u^*_{\rm{s}}\}$ and $u^*_{\rm{s}} < x^{\NE}_{\rm{a}}$, we have that 
\begin{align}
    \nonumber
    f_{C, \rm{s}}(u^{\NE}) &<
    f_{C, \rm{s}}(u^*_{\rm{s}})
    \\ \nonumber &= 
    f_{D, \rm{s}}(u^*_{\rm{s}})
    \\ \nonumber &<
    f_{D, \rm{s}}(u^{\NE}).
\end{align}
Now, $u^{\NE}_{\rm{s}} = u^*_{\rm{s}} < x^{\NE}_{\rm{a}} = u^{\NE}$, and $x^{\NE}_{\rm{a}} \leq p_{\rm{a}} < u^*_{\rm{a}}$, so $u^{\NE}_{\rm{s}} < u^{\NE} < u^*_{\rm{a}}$.
Hence $W(u^{\NE}_{\rm{s}}) \leq W(u^{\NE})$. \hfill $\blacksquare$\\

\noindent Finally, we include the proof of Proposition~\ref{prop:PD_characterization}.

\textit{Proof of Proposition~\ref{prop:PD_characterization}:} Since $S < P$ and $R < T$, defecting is a dominant strategy for selfish agents, i.e. $f_{D, \rm{s}}(u) > f_{C, \rm{s}}(u)$ for all $u \in [0,1]$.
Thus, in any Nash equilibrium, $x^{\NE}_{\rm{s}} = 0$.
Next, we identify the values of $x^{\NE}_{\rm{a}} \in [0,p_{\rm{a}}]$ that result in a Nash equilibrium for altruists.
In particular, $x^{\NE}_{\rm{a}}$ must satisfy one of the Nash equilibrium conditions in~\eqref{eq:equilibrium_conditions}.
Hence, the social welfare at equilibrium in a Prisoner's Dilemma is characterized by
\begin{equation}
    W(u^{\NE}) =
    \begin{cases}
        P & \text{if } x^{\NE}_{\rm{a}} = 0\\
        P - \frac{\beta^2}{4\delta} & \text{if } x^{\NE}_{\rm{a}} = u^*_{\rm{a}} \\
        \delta p^2_{\rm{a}} + \beta p_{\rm{a}} + P & \text{if } x^{\NE}_{\rm{a}}  = p_{\rm{a}}
    \end{cases}.
\end{equation}

First, assume $\delta = 0$.
Then $f_{C, \rm{a}}(u) = S+T - 2P > 0$ (since $R > P$), so $x^{\NE}_{\rm{a}} = p_{\rm{a}}$ is the only Nash equilibrium:
\begin{equation}
    \PI(\PD(p_{\rm{a}})) = \frac{\delta p_{\rm{a}}^2 + \beta p_{\rm{a}}}{P} + 1.
\end{equation}
Since $\beta > 0$ and $\delta = 0$, $\PI(\PD(p_{\rm{a}})) \geq 1$.

We next consider $\delta > 0$.
Then $u^*_{\rm{a}} < 1$, since $R > P$.
Now, if $u^*_{\rm{a}} \leq 0$, then $x^{\NE}_{\rm{a}} = 0$ and so $\PI(\PD(p_{\rm{a}})) = 1$.
Hence we can just consider $u^*_{\rm{a}} \in (0,1)$.
Then, social welfare attains the global minimum value of
\begin{equation}\label{eq:global_min}
    W(u^*_{\rm{a}}) = P - \frac{(2P-(S+T))^2}{4(R+P-(S+T))},
\end{equation}
and attains a local maximum value of $W(1) = R$ (since \mbox{$R > P$}).
It holds that $f_{D, \rm{a}}(0) = 2P-(S+T) > 0$, and recall $f_{C, \rm{a}}(u) = -f_{D, \rm{a}}(u)$.
Thus, $x^{\NE} = (0, 0)$ is a Nash equilibrium for any $p_{\rm{a}} \leq u^*_{\rm{a}}$.
Now, because the payoff functions are affine, the only equilibrium $x^{\NE}_{\rm{a}} \in (0, 1)$ for which $f_{C, \rm{a}}(x^{\NE}_{\rm{a}}) = f_{C, \rm{a}}(x^{\NE}_{\rm{a}})$ is $u^*_{\rm{a}}$.
Therefore, if $u^*_{\rm{a}} \leq p_{\rm{a}}$, then $\xVector^{\NE} = (u^*_{\rm{a}}, 0)$.
Further, $x^{\NE}_{\rm{a}} = p_{\rm{a}}$ (i.e. $p_{\rm{a}} \leq u^*_{\rm{a}}$) if and only if $f_{C, \rm{a}}(p_{\rm{a}}) \geq f_{D, \rm{a}}(p_{\rm{a}})$.
Hence, for $\delta > 0$, the set of Nash equilibria for $\PD(p_{\rm{a}})$ is summarized by $x^{\NE}_{\rm{a}}$ as follows:
\begin{align}
    x^{\NE}_{\rm{a}} =
    \begin{cases} 
        0, & \text{ if } p_{\rm{a}} < u^*_{\rm{a}} \\
        \{ 0, u^*_{\rm{a}}, p_{\rm{a}}\}, & \text{ if } p_{\rm{a}} \geq u^*_{\rm{a}}
    \end{cases}.
\end{align}
Thus, when $W$ is strictly convex, the resulting perversity index given by~\eqref{eq:PD_convex_PI_characterization} is obtained.
To see that $\PI(\PD(p_{\rm{a}})) \leq 1$, notice that $\frac{\beta^2}{4P\delta} \geq 0$, since $\delta > 0$ and $\beta^2 \geq 0$.

Finally, we consider when $W$ is strictly concave ($\delta < 0$).
It can be shown $W$ attains the global maximum value of
\begin{equation}\label{eq:global_max}
    W(u^*_{\rm{a}}) = P - \frac{(2P-(S+T))^2}{4(R+P-(S+T))},
\end{equation}
and local minimum value of $W(0) = P$.
Now, we need to identify the values $x^{\NE}_{\rm{a}}$ can attain.
Since $x^{\NE}_{\rm{s}} = 0$, and by Lemma~\ref{lem:X_tau_cardinality}, we know that 
\begin{align}
    x^{\NE}_{\rm{a}} =
    \begin{cases} 
        p_{\rm{a}}, & \text{ if } p_{\rm{a}} < u^*_{\rm{a}} \\
        u^*_{\rm{a}}, & \text{ if } p_{\rm{a}} \geq u^*_{\rm{a}}
    \end{cases}.
\end{align}
Thus, when $W$ is strictly concave, the perversity index given by~\eqref{eq:PD_concave_PI_characterization} is obtained.
To see that $\PI(\PD(p_{\rm{a}})) \geq 1$, notice that since $\delta < 0$ and $\beta^2 \geq 0$, it follows that $\frac{\beta^2}{4P\delta} \leq 0$. \hfill $\blacksquare$



\end{document}